\documentclass[%
 aip,
 jcp,
 amsmath,amssymb,
 reprint%
]{revtex4-1}

\usepackage{graphicx}
\usepackage{dcolumn}
\usepackage{bm}

\usepackage[utf8]{inputenc}
\usepackage[T1]{fontenc}
\usepackage{mathptmx}
\usepackage{graphicx}
\usepackage{subcaption}
\usepackage{afterpage}
\usepackage{comment}
\usepackage{siunitx}
\usepackage{cleveref}
\usepackage{float}

\begin{document}


\title{An Entropy-Maximization Approach to Automated Training Set Generation for Interatomic Potentials}

\author{Mariia Karabin}
\affiliation{Department of Chemistry, Clemson University, Clemson, SC 29634 USA}
\affiliation{Theoretical Division T-1, Los Alamos National Laboratory, Los Alamos, NM 87545 USA}
\author{Danny Perez}
\email[]{danny_perez@lanl.gov}
\affiliation{Theoretical Division T-1, Los Alamos National Laboratory, Los Alamos, NM 87545 USA}

\date{\today}

\begin{abstract}
Machine learning (ML)-based interatomic potentials are currently garnering a lot of attention as they strive to achieve the accuracy of electronic structure methods at the computational cost of empirical potentials. Given their generic functional forms, the transferability of these potentials is highly dependent on the quality of the training set, the generation of which is a highly labor-intensive activity. Good training sets should at once contain a very diverse set of configurations while avoiding redundancies that incur cost without providing benefits.
We formalize these requirements in a local entropy maximization framework and propose an automated sampling scheme to sample from this objective function. We show that this approach generates much more diverse training sets than unbiased sampling and is competitive with hand-crafted training sets. 
\end{abstract}

\maketitle

\section{\label{sec:level1}Introduction}

The practical usefulness of atomistic simulations ultimately relies on the availability of interatomic potentials that are able to provide reliable energies and forces at a sufficiently affordable computational cost. Since electronic structure calculations using techniques such as density functional theory (DFT) are often prohibitively expensive, simplified empirical forms have been the norm, 
especially for molecular dynamical applications. Early empirical potentials were traditionally highly computationally efficient but often lacked in accuracy and transferability. 

Over the last few years, the need to bridge the gap between empirical methods and direct electronic structure calculations has driven the explosive development of machine learning (ML) based approaches that aim to combine the accuracy of the electronic structure methods and the efficiency of the early simplified potentials. The two main components of the ML-based potentials are the representation of the atomic structures with a set of generic descriptors that characterize local atomic environments
and the use of large amounts of data to train complex non-linear functions of the descriptors to reproduce reference electronic structure calculations (energies, forces, stresses, etc.). 

While ML-based potentials have proved able to capture subtle features of the training data, their ability to extrapolate to situations that markedly differ from those encountered during training remains limited\cite{PhysRevB.99.184305}. Therefore, the accuracy of ML-based potentials is highly dependent on the choice of the training set, which should i) cover as much of the relevant configuration space as possible, and ii) remain sufficiently compact so that the cost of computing the reference values with quantum calculations and training the model remains affordable. Traditionally, training set generation has been a highly labor-intensive activity that relies on  physical intuition in order to select the  configurations that should be included.  

Different approaches have been proposed to address the first objective using sampling strategies \cite{der2018,der18,PhysRevB.99.064114,haj17,Chmiela2018,LEE201995}, including evolutionary structural searches \cite{chan19}, normal mode sampling \cite{C6SC05720A}, and exploration of the potential energy surface using on-the-fly approximations of the target potential \cite{der2018}. Training set configurations are also selected from DFT-MD simulations\cite{jeo2018}. 

The second objective is often achieved by sub-sampling larger data sets. Possible approaches include random selection \cite{doi:10.1021/acs.jpcc.6b10908}, binning-based sub-sampling to achieve uniform representation of relevant quantities like atomic forces \cite{Huan2017}, clustering in descriptors space to identify distinct groups \cite{Huan2017}. Finally, a number of recent approaches incrementally include data to the training set based on whether the prediction of the properties of new configurations require extrapolation \cite{Behler_2014,PhysRevLett.93.165501,PODRYABINKIN2017171,PhysRevB.99.064114,GUBAEV2019148}. These approaches differ by the algorithm type and query strategy\cite{PODRYABINKIN2017171,PhysRevB.99.064114}.

In this manuscript, we unify the diversity and non-redundancy objectives in a simple local approach where the diversity of atomic environment within individual configurations (as measured by an entropy metric) is maximized subject to the constraint that the configurations do not contain unphysical configurations (i.e., overlapping atoms). This objective is embodied in a  generic effective potential energy function whose low-lying local minima are good candidates for inclusion in a training set. Such minima are sampled using a simple annealing scheme that can be easily automated. Importantly, this effective energy is not meant as an approximation to the energy of the actual target system; instead it is an abstract construct that enables the creation of material-agnostic training sets. In this sense, our approach aims at creating a "universal" set of configurations that captures a very wide range of local environments and does not focus solely on low-lying energy structures. The large volume of configuration space covered entails a tradeoff between the size of the training set and the target accuracy, but the high transferability it affords is important to capture high-energy, far from equilibrium effects that can occur in extreme conditions, such as under irradiation or at high pressures. A global approach where diversity maximization is carried out globally over the whole training set is currently in development and will be reported in an upcoming manuscript.



\section{Methods}

\subsubsection{Entropy maximization approach}

Implementing these ideas in practice requires first defining a set of atomic descriptors $\{\bf{q}\}$ that characterize the local environment of each atom, and then defining a measure of the diversity of the distribution of these descriptors within a configuration of atoms. A wide array of atomic descriptors have been proposed in the literature\cite{doi:10.1021/acs.jctc.8b00110}, as these form the inputs of many machine learning approaches that learn atomic energies. The method we propose is agnostic to the specific choice of descriptors so as long as they are differentiable functions of atomic positions. In the following, the set of $m$ descriptors $q_{i,k}$ of the local atomic environment of atom $i$ is arranged into a vector ${\bf q}_i$ of length $m$.
As a measure of the diversity, we use the entropy of the $m$-dimensional distribution of atomic descriptors $S(\{\bf{q}\})$ contained in a given configuration of atoms, which is a natural choice in this case: it is maximized for a uniform distribution of descriptors and minimized for configurations where all environments are identical, i.e., it promotes diversity and penalizes redundancy. The effective energy we propose is therefore of the form:

\begin{eqnarray} \label{eq:effective_energy}
    V &=& E_\mathrm{repulsive} -K S(\{\bf{q}\})
    \end{eqnarray}

where $E_\mathrm{repulsive}$ is a short range repulsive term that penalizes very short distances between atoms (so as to enforce an excluded volume around each atom) and $K$ is a so-called entropy scaling coefficient that tunes the relative importance of the entropy and of the repulsive contribution. Local minima of this function can therefore be expected to contain a high diversity of different environments without any two atoms being unphysically close. We postulate that low-lying minima of this effective potentials are therefore good targets for inclusion in a training set.

A number of approaches have been proposed to numerically estimate the entropy of a distribution of descriptors. In the following, we adopt a simple nonparametric form where the local density is approximated using the first neighbor distance (in descriptor space)\cite{sztaki1417}. In this case, the estimator is of the form:

\begin{eqnarray} \label{eq:2}
    S(\{q\}) = \frac{1}{n} \sum_{j=1}^n \ln (n \cdot \Delta q^\mathrm{min}_j)
\end{eqnarray} 
where $\Delta q^\mathrm{min}_j$ is the nearest-neighbor distance in descriptor space, i.e., it is 
the minimal distance between the descriptor of atom $j$ and those of any other atom in the configuration, i.e.,

\begin{eqnarray} \label{eq:3}
    \Delta q^\mathrm{min}_j &=& \min_{l} \sqrt{ \Delta {\bf q}_{jl} \cdot  \Delta {\bf q}_{jl}  }.
\end{eqnarray} 
and $n$ is the number of atoms in the cell. This specific choice is not expected to be critical and other estimators could be used instead.

\subsubsection{Computational details}

\begin{figure}
\includegraphics[width=\linewidth]{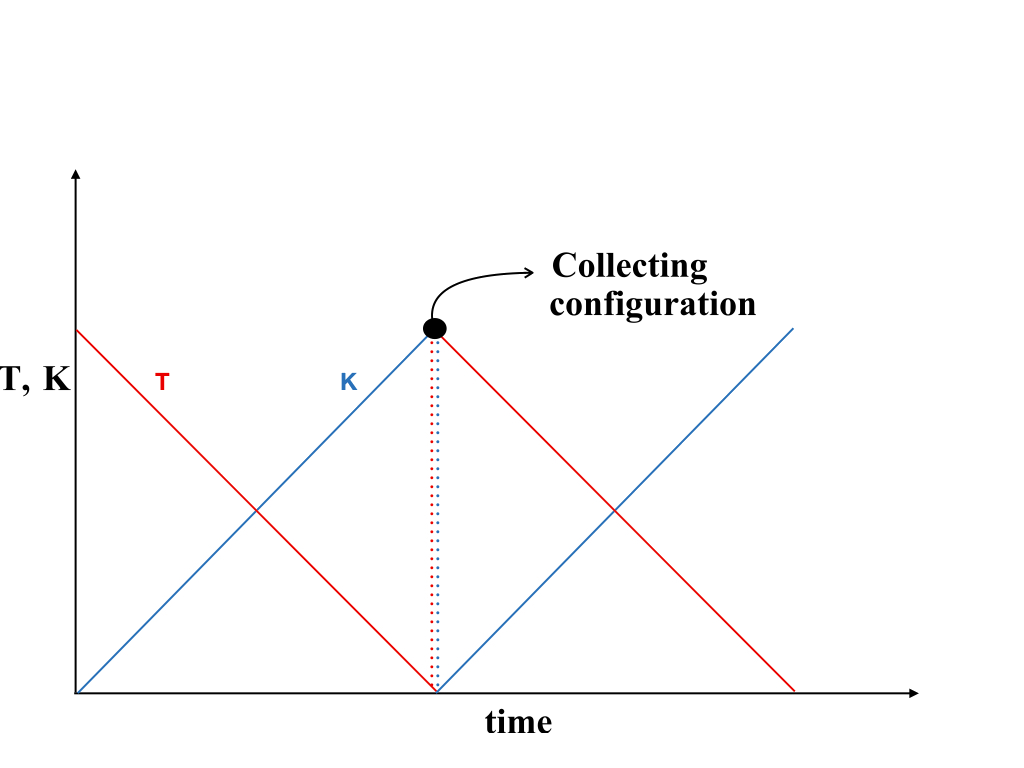}
\caption{\label{fig:anneal} Schematic representation of the cyclic annealing procedure. See text for details.}
\end{figure}

The training set is incrementally constructed by adding independent local minima of the effective energy Eq.\ \ref{eq:effective_energy}. As the effective potential (much like actual potentials) is very rough, a simple annealing procedure was introduced, as illustrated in Fig.\ref{fig:anneal}. Note that the aim is not to locate the global minimum of the effective energy but simply to avoid trapping in low entropy configurations. Our annealing procedure proceeds through a simultaneous ramping down of the temperature and ramping up of $K$. The goal is to initially favor a thorough shuffling of the atomic positions and avoid correlations between successive configurations by using a high temperature ($10,000$K) and no entropy bias. Entropy maximization is then gradually favored by linearly decreasing the temperature down to 0 and ramping up $K$ to $1000$ eV. The resulting configuration is then harvested and added to the training set. The cycle then simply repeats.

\begin{figure}
\includegraphics[width=\linewidth]{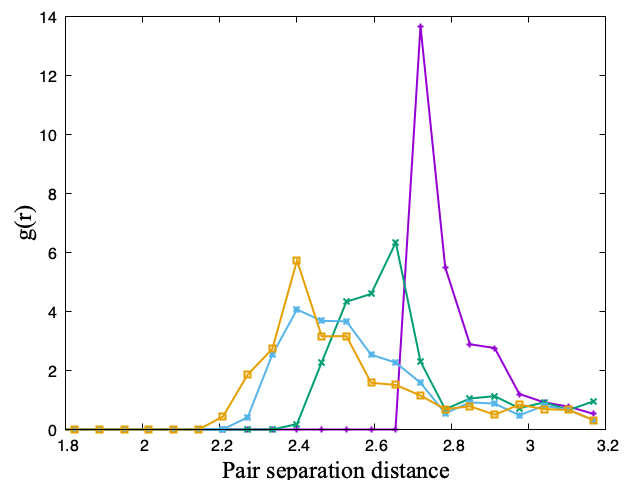}
\caption{\label{fig:rdf} Radial distribution function for an annealed configuration obtained with $K=0$ eV (purple), $K=1000$ eV (green), $K=2000$ eV (blue), and $K=3000$ eV (yellow).}
\end{figure}

The maximal value of entropy scaling factor $K$ was tuned so as not to overwhelm $E_\mathrm{repulsive}$ while still providing a strong driving force for the maximization of the entropy. As shown in Fig.\ \ref{fig:rdf}, increasing too far $K$ yields configurations where some pairs of atoms become separated by very short distances. Large values of $K$ also yield stiff effective potentials that are prone to instabilities during the MD annealing. We therefore settled on a maximal value of $K=1000$ eV. Note that this choice depends on the number of atoms in the simulation cell, as the entropy so-defined is intensive, but the repulsive contribution is extensive. 

The spatial scale of the problem was chosen to be representative of tungsten atoms, but the training set is fully generic, and can therefore be rescaled as needed to describe other elements. In the following, we used cells containing $n=39$ atoms with a volume of $9.54$x$9.54$x$14.31$ \AA. This corresponds to a density of $0.03$ atoms/\AA$^3$, as compared to a bulk BCC density of $0.062$ atoms/\AA$^3$ for tungsten. 
The number of atoms was chosen so that cubic-scaling DFT calculations would be affordable whereas the volume was chosen so that both high and low density regions could coexist within the same simulation cell, thereby creating configurations that contain to bulk, surfaces, and voids. While even larger volumes allowed for larger descriptor entropies because of additional opportunities to create complex atomic arrangements, local minima of the effective energy at low density tend to contain high proportion of 1D filament-like structures and of gas-like configurations. If such configurations are deemed relevant, a training set can constructed by combining a range of different cell sizes. This possibility will be explored a future study.

In the following, atomic environments were described in terms of the so-called bispectrum components originally developed in the context of the Gaussian Approximation Potentials (GAP) potentials\cite{PhysRevLett.104.136403}, and then adopted by the SNAP approach \cite{2015JCoPh.285..316T}. These descriptors are invariants of an expansion of the the density of neighboring atoms around a central atom in terms of hyperspherical harmonics.
They are attractive because they are rotationally and permutationally invariant, which facilitates the development of energy expressions that inherit from these same properties. Progressively higher-order components then capture increasingly fine details of the distributions of neighboring atoms. Details of the computation of the bispectrum components can be found in the original publications\cite{PhysRevLett.104.136403}. The results presented below used the first 6 bispectrum components to characterize each atomic environment.




In the following, $E_\mathrm{repulsive}$ follows the form proposed by Clarke and Smith\cite{cla86}:

\begin{eqnarray} \label{eq:5}
E_\mathrm{repulsive} &=& \sum_i \sum_j \frac{E_0}{n-m} \left[m \left(\frac{r_0}{r_{ij}} \right)^n -  n \left(\frac{r_0}{r_{ij}} \right)^m\right] 
\end{eqnarray}
with $E_0 = 1$ eV, $n = 8$, $r_0 = 2.7$ \AA,  and $m=4$.  The potential was truncated at $r= 2.71$\AA, and shifted to zero at the cutoff so as to capture only the repulsive part of the potential. The results are not expected to be sensitive to the specific form of the repulsive potential, as its only purpose is to enforce excluded volumes around each atom.



\section{Results}


\subsubsection{Characterization of the descriptor diversity}

\begin{figure} 
  \begin{subfigure}[b]{0.57\linewidth}
    \centering
    \includegraphics[width=0.97\linewidth]{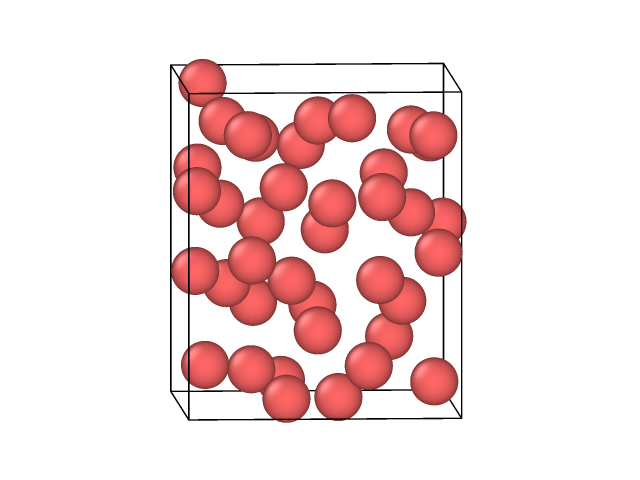} 
    \label{fig:a} 
    \vspace{-1ex}
  \end{subfigure}\hspace{-0.55in}
  \begin{subfigure}[b]{0.57\linewidth}
    \centering
    \includegraphics[width=0.97\linewidth]{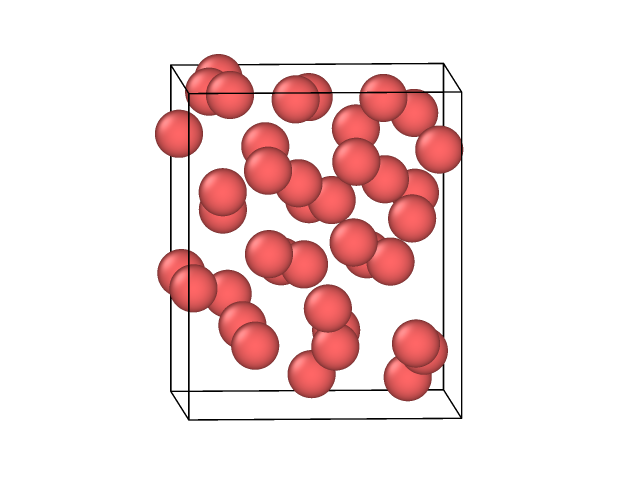} 
    \label{fig:b} 
    \vspace{-1ex}
  \end{subfigure}
  \caption{Configurations generated with the entropy maximization approach}
  \label{fig:config} 
\end{figure}

10,000 configurations of $n=39$ atoms were generated using the procedure described above. A few representative configurations are shown in Fig. \ref{fig:config}. The ensemble of these configurations is referred to as the "biased" dataset. As a point of comparison, we compare the results with a so-called "unbiased" reference dataset, where configurations were sampled from an MD simulation at a temperature of $T=10,000$K with $K=0$, i.e., without attempting to maximize the entropy but while enforcing excluded volume constraints. As expected, the average descriptor entropy of configurations in the biased set ($S\sim 4.4$) is larger than that of the unbiased set ($S\sim 3.2$). 

\begin{figure}[!htb]
\includegraphics[scale=0.55]{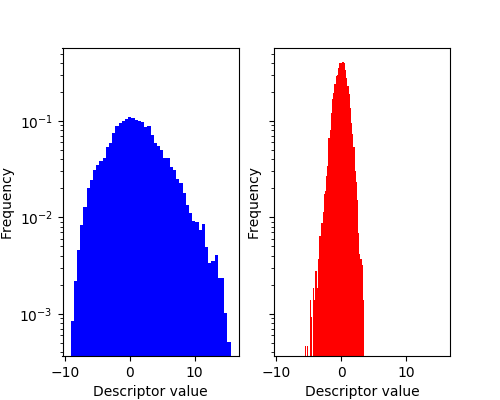}
\caption{\label{fig:hist0-1} Distribution of the first descriptor. Left: biased dataset; Right: unbiased dataset.}
\end{figure}

The consequences of this increase in entropy can be appreciated by contrasting the distribution of individual descriptors over the biased and unbiased sets, as shown in Fig.\ \ref{fig:hist0-1} for the first bispectrum component. The distribution over the biased set is clearly much broader than its unbiased counterpart, which was the intended behavior. This shows that, even if the entropy maximization was applied locally to each configuration, the procedure yields  broad distributions over the whole training set. Perhaps surprisingly,
computing high ($>6$th) order descriptors shows that their distribution is also broadened in the biased set. A multiple correlation analysis demonstrates that this results from the fact that the descriptors are not mutually linearly independent; on average, we observe a correlation coefficient of about 0.7 for high-order descriptors against the first 6.


\begin{figure}[!htb]
\includegraphics[scale=0.55]{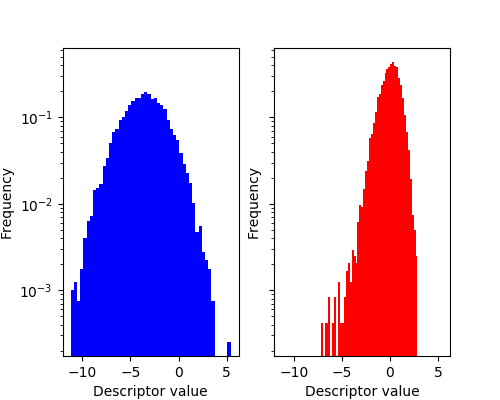}
\caption{\label{fig:hist0-2} Distribution of the eight descriptor. Left: biased dataset; Right: unbiased dataset.}
\end{figure}

\subsubsection{Error estimations on trained potentials: biased vs unbiased datasets}

In order to quantify the quality of the generated training set, we consider an ML scenario where energies and forces are computed through a gaussian process regression (GPR) parameterized on a given training set \cite{Ras06,han14} as was done in the GAP approach \cite{PhysRevB.90.104108,PhysRevB.99.184305}. We consider the GPR to act as an interpolator that exactly reproduces reference data at training points. In this setting, the distance to the nearest training point (in descriptor space) is a simple surrogate for the error in predictions at test points. The shift and scale transformation that renders the distribution of each descriptors in the unbiased dataset mean free and unit variance was applied to both training and testing sets in order to uniformize the scales of each descriptors. In the following, the training sets contains 6000 randomly selected atomic environments, and testing sets 3000. Results were averaged over 1000 random decompositions between testing and training sets. Distances were measured in the 6-dimensional space spanned by the biased descriptors.

This metric is used to first compare the quality of the unbiased and biased datasets. Table\ \ref{tab:errors} shows the mean distances obtained using different combinations of training and testing sets. Training on the unbiased set performs well (i.e., the mean error is low) when testing points are also sampled from the unbiased set. The distances however become very large when testing points are sampled from the biased set. This is a reflection of the fact that the distribution of descriptors in the unbiased set has a relatively narrow support: the training distribution is therefore dense over the support (yielding small distances when the test points fall within the support) but very large distances when the test points fall outside of the support (i.e., when test points are sampled from the biased distribution). In this latter case, the GPR is extrapolating, leading to large distances. This illustrates an important tradeoff: a potential trained on a narrow set of configurations can be expected to do well when used on configurations close to this narrow set; it will however do poorly when departing from it. In contrast, the distances obtained from training on the biased set show little dependence on the nature of the testing set, as the GPR is not forced to extrapolate outside of the support of the training set. Another tradeoff apparent here is that the mean distance when training on the biased set and testing on the unbiased set are higher than those observed when training and testing on the unbiased set. This follows from the inverse relationship between the size of the support and the density of points in descriptor space at fixed number of training points. If one is aiming at transferability, the biased training set is however clearly favorable to the unbiased one.

\subsubsection{Error estimations on trained potentials: biased vs hand-crafted datasets}

A more stringent test of our approach is to compare the biased training set with an "hand-crafted" dataset that was used to train potentials reported in the literature. To this end, we select a well established training set that was used in the development of a number of recent potentials for tungsten\cite{PhysRevB.99.184305}.

\begin{figure}[!htb]
\includegraphics[scale=0.55]{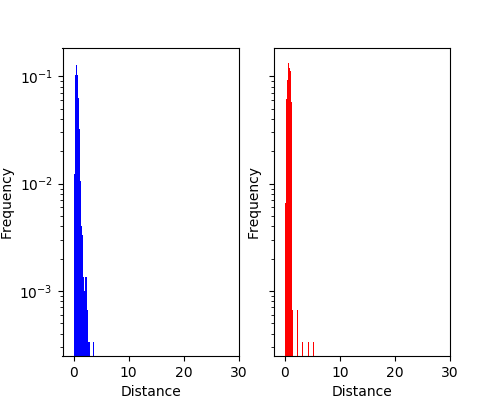}
\includegraphics[scale=0.55]{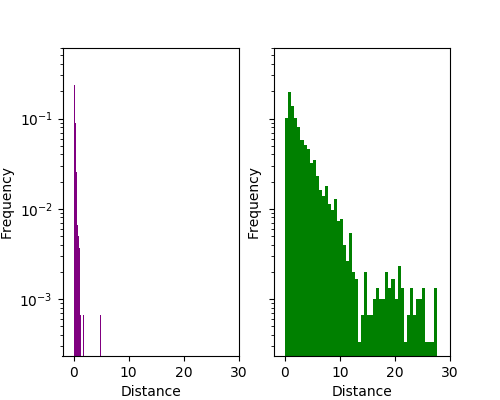}
\caption{\label{fig:dist} Distribution of the distances to the closest training point for different combinations of training and testing sets. Training from the biased set and testing from the hand-crafted set (red); training and testing from the biased sets (blue); training and testing from the hand-crafted set (purple); training from the hand-crafted set and testing from the biased set (green).}
\end{figure}

Table\ \ref{tab:errors} reports the results of the distances to the nearest training point for training and testing sets drawn from the hand-crafted and biased sets. The results are largely similar to that observed for the testing set. Training and testing from the hand-crafted set yields low errors as the support of both training and testing distributions is very narrow (c.f. purple histogram in Fig.\ \ref{fig:dist}), but these increase dramatically upon switching the test set to the unbiased set, again because extrapolation is then required (c.f. the very long tail in the green distribution in Fig.\ \ref{fig:dist}). In contrast, training from the biased set yields results that are similar for both testing sets (c.f. blue and red histograms in Fig.\ \ref{fig:dist}). These results suggest that the automated training set should be competitive with the hand-crafted set as it contains a more diverse distribution of atomic environments.

\begin{figure}[!htb]
\includegraphics[scale=0.55]{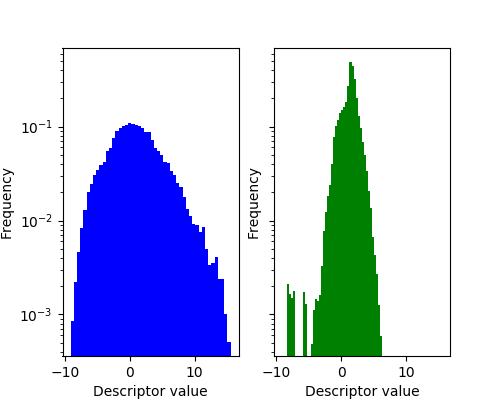}
\caption{\label{fig:hist1-1} Distribution of the first descriptor. Left: biased dataset; Right: hand-crafted dataset.}
\end{figure}
	
\begin{figure}[!htb]
\includegraphics[scale=0.55]{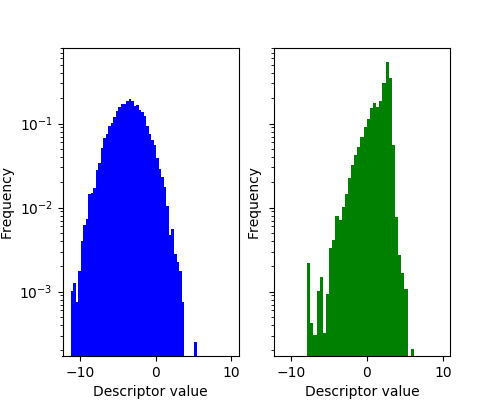}
\caption{\label{fig:hist1-2} Distribution of the eight descriptor. Left: biased dataset; Right: hand-crafted dataset.}
\end{figure}

Close analysis reveals that this characterization comes with caveats. Indeed, as shown in Fig.\ \ref{fig:hist1-1}, the distribution of descriptors is in general significantly wider in the biased set than in the hand-crafted set. However, the distribution of some descriptors in the hand-crafted set is strongly peaked, as it contains a high proportion of crystalline local environments. In some cases, the peak falls into a region where the density descriptors in the biased dataset is low, c.f. Fig.\ \ref{fig:hist1-2}, which can limit the accuracy of predictions carried out using the biased set alone for training.
This translates into an increasing errors when testing with the hand-crafted states 
as the space of descriptors in which the GPR interpolation is carried out increases to tens or hundreds of dimensions. Note however that even in this case, the errors remain below that of training with the hand-crafted set and testing with the biased set. This limitation could potentially be addressed by increasing the size of the biased space or by explicitly favoring high-symmetry local order. 

This observation illustrates the tradeoffs discussed above: if one seeks an highly accurate potentials that is valid in a small region of the possible configuration space of the problem (e.g. in BCC crystalline configurations), a narrow but tailored training set is likely to perform better; on the other hand, if transferability is paramount, an approach that explicitly favors diversity as the one proposed here is highly beneficial. In practice, these two extremes can be bridged to achieve both high accuracy in low-energy states and transferability to higher-energy configurations.

\begin{table}
\caption{\label{tab:errors}Error estimations on trained potentials: biased vs unbiased datasets}
\begin{ruledtabular}
\begin{tabular}{ccddd}
Training set&Testing set&\mbox{Mean}&\mbox{Median}&\mbox{Max}\\
\hline
Unbiased&Unbiased&\mbox{0.2294}&\mbox{0.1827}&\mbox{5.9339}\\
Unbiased&Biased&\mbox{4.6949}&\mbox{3.5311}&\mbox{32.0983}\\
Biased&Unbiased&\mbox{0.9053}&\mbox{0.9068}&\mbox{2.3402}\\
Biased&Biased&\mbox{0.8541}&\mbox{0.7846}&\mbox{5.7246}\\
\end{tabular}
\end{ruledtabular}
\end{table}

\begin{table}
\caption{\label{tab:errors2}Error estimations on trained potentials: biased vs hand-crafted datasets}
\begin{ruledtabular}
\begin{tabular}{ccddd}
Training set&Testing set&\mbox{Mean}&\mbox{Median}&\mbox{Max}\\
\hline
Hand-crafted&Hand-crafted&\mbox{0.2263}&\mbox{0.1562}&\mbox{7.6514}\\
Hand-crafted&Biased&\mbox{3.6680}&\mbox{2.3153}&\mbox{36.4918}\\
Biased&Hand-crafted&\mbox{0.9923}&\mbox{1.0014}&\mbox{7.9049}\\
Biased&Biased&\mbox{0.8541}&\mbox{0.7846}&\mbox{5.7246}\\
\end{tabular}
\end{ruledtabular}
\end{table}

\section{Conclusions}

We introduced a sampling-based approach for the automated training set generation of interatomic potentials.
Configurations are generated by sampling low-lying minima of an effective potential energy function that explicitly favors the diversity of the local atomic environment through an entropy maximization process. The generated training set is shown to be more diverse than that generated by a random sampling procedure and even compared to hand-crafted sets used in state-of-the-art machine-learned potentials, which promises improved transferability. Extensions to global entropy-maximization over the whole training set (in contrast to the local configuration-by-configuration optimization presented here) is in development and will be reported in an upcoming publication.

\section{Acknowledgements}
We thank Mitchell Wood and Nicholas Lubbers for stimulating discussions.
This research was supported by the Exascale Computing Project (17-SC-20-SC), a collaborative effort of the U.S. Department of Energy Office of Science and the National Nuclear Security Administration. Los Alamos National Laboratory is operated by Triad National Security LLC, for the National Nuclear Security administration of the U.S. DOE under Contract No. 89233218CNA0000001.

\nocite{*}
\bibliography{aipsamp}

\end{document}